%% file: JournalBridging_Cybernatics.tex
\documentclass[journal]{IEEEtran}
\usepackage{setspace}
\usepackage{amstext}
\usepackage{cite}
\usepackage{amsmath}
\usepackage{amsbsy}
\usepackage{amssymb}
\usepackage{mathtools}
\usepackage{graphicx}
\usepackage{caption}
\usepackage{subcaption}
\usepackage{multirow}
\usepackage{booktabs}
\input{jamesmacros.tex}

\usepackage{soul}

\begin{document}

\pagenumbering{arabic}
\title{
Bayesian Intent Prediction in Object Tracking Using Bridging Distributions}
\author{Bashar I. Ahmad$^\dagger$, 
        James K. Murphy$^\dagger$,  
       Patrick M. Langdon and
        Simon J. Godsill 
\thanks{B. I. Ahmad*, J. K. Murphy, P. M. Langdon, S. J. Godsill are with the Engineering Department, University of Cambridge, Trumpington Street, Cambridge, UK, CB2 1PZ. Emails:\{bia23, jm362, pml24, sjg30\}@cam.ac.uk.}
\thanks{$^\dagger$The first two authors made an equal contribution to this work.}
}
\markboth{Submitted to IEEE Transactions on CYBERNETICS,~Vol.~XX, No.~XX, XXXX~XXXX}%
{Shell \MakeLowercase{\textit{et al.}}: Bare Demo of IEEEtran.cls for Journals}

\maketitle 

\begin{abstract}
In several application areas, such as human computer interaction,  surveillance and defence, determining the  intent of a tracked object  enables  systems to aid the user/operator and facilitate effective, possibly automated, decision making. In this paper, we propose a  probabilistic inference approach  that permits the prediction, well in advance, of the intended destination of a tracked object and its future trajectory. Within the framework introduced here, the observed partial track of the object is modeled as  being part of a   Markov bridge terminating at  its destination, since the target path, albeit random, must end at  the intended endpoint. This captures the underlying long term dependencies in the  trajectory, as dictated by the object intent.  By determining the likelihood of the partial track being drawn from a particular constructed bridge, the probability of each of a number of possible  destinations   is evaluated. These bridges can also be  employed to produce refined estimates of the latent system state (e.g. object position, velocity, etc.), predict  its future values (up until reaching the designated endpoint) and estimate the time of arrival.   This is shown to lead to a low complexity Kalman-filter-based implementation of the inference routine,   where any linear Gaussian motion model, including  the destination reverting ones, can be applied. Free hand pointing gestures  data  collected in  an instrumented vehicle and synthetic trajectories of a  vessel  heading towards multiple possible harbours  are utilised to demonstrate the effectiveness of the proposed   approach. 
\end{abstract}
\begin{IEEEkeywords}
Bayesian inference, Kalman filtering, tracking, maritime surveillance,  Human computer interactions.
\end{IEEEkeywords}
\section{Introduction}
\IEEEPARstart{I}n several application areas such as surveillance, defence and human computer interaction (HCI), the trajectory of a tracked object, e.g.  a vessel, jet, pedestrian   or pointing apparatus, is driven by its final destination. Thus,   \textit{a priori} knowledge of the object endpoint can not only offer vital information on  intent, unveil potential conflict or threat and enable task facilitation strategies,  but can also produce more accurate tracking routines  \cite{castanon1985algorithms,ristic2008statistical,baccarelli1998recursive,piciarelli2008trajectory,wang2011intent,fanaswala2013detection,nunez2013hard,pallotta2013vessel,pallotta2014context}. In this paper, we address the problem of predicting the intended destination of a tracked object from a finite set of possible endpoints and  the future values of its hidden \textit{state} (e.g.  position, velocity, etc.), given the available   noisy observations. This can be viewed as a means to assist or automate timely decision making, planning and resources allocation  at a higher system level, compared with a conventional sensor-level tracker. The latter typically  focuses on inferring the current value of the  latent \textit{state} $X_t$, with several well-established  algorithms \cite{li2003survey,li2005survey,BarShalomBook2011}. 

To motivate the work presented here, consider the following two examples:
\\ \textit{1)  Maritime Surveillance:} analysing the route and determining the destination of vessels in a given geographical area is necessary for maintaining maritime situational awareness (MSA), critical for maritime safety. This permits  identification of potential threats, opportunities or malicious behaviour, allowing the protection of assets or other reactive actions  \cite{castanon1985algorithms,ristic2008statistical,pallotta2013vessel,pallotta2014context}. Given the complex nature of typical maritime traffic as well as the vast amounts of available information on tracked targets, there is a growing interest in  increasing the degree of automation in MSA\ systems by unveiling the intent of object(s) of interest from available low level tracking data.   Thus, there is a notable demand for low-complexity and reliable destination and  trajectory prediction techniques. Similar challenges can be found in aerospace surveillance, with aircraft in lieu of vessels.    
 \\ \textit{2) Interacting with touchscreen: } touchscreens  are becoming an integrated part of modern vehicles due to their ability to present large quantities of  in-vehicle infotainment system data and offering additional design flexibility through a combined display-input–-feedback module \cite{harvey2013usability,burnett2001ubiquitous}.  Using these displays entails undertaking a free hand pointing gesture and dedicating a considerable amount of attention  that would otherwise be available for driving.  Hence, such interactions can act as a distractor from the primary task of driving and have  safety implications \cite{klauer2006impact}. The early inference of the intended on-screen item  of the free hand pointing gesture can simplify and expedite  the selection task, thus significantly improving the usability of   in-vehicle touchscreens by reducing  distractions (see  \cite{ahmadCyberTrans2015} for an overview of  the intent-aware display concept).

There is a wide range of other applications that can benefit from knowing the intent of a target of interest.  For example,  predicting the destination of an intruder in a  perimeter, inferring the future position of a moving person in a given area to take appropriate robotic actions \cite{ziebart2009planning}, and advanced driver assistance system in vehicles \cite{bando2013unsupervised}, to name a few.   
\subsection{Contributions}
The main contribution of this paper is the development  of  a simple  and robust Bayesian intent inference approach that models the available partial trajectory of a tracked object  as a  random bridge terminating at a nominal destination. It can employ any  linear Gaussian motion model, including \emph{Linear Destination Reverting} (LDR) models, such as those detailed in \ssections{MRD}{ERV}, which are intrinsically driven by the  endpoint  of the tracked object (they were used in \cite{ahmadCyberTrans2015}). The  bridging framework introduced here capitalises on the premise that the path of the object, albeit random, must end at  the intended destination.  Since the endpoint is unknown \textit{a priori}, a bridge for each possible destination is constructed. This encapsulates the long term dependencies in the object trajectory due to  premeditated actions guided by  intent.
By determining the likelihood of the observed  partial    track being drawn from a particular bridge, the probability of each  nominal endpoint, along with a refined estimate of the current object state $X_t$ and its  future values $X_s$ (for $s>t$), can be evaluated. This is accomplished via Kalman-filter-based inference, amenable to parallelisation.  

Notably,  the proposed approach in this paper does not impose  prior knowledge of the arrival time $T$ at the intended destination $D$. A conservatively chosen prior distribution of the possible times of arrival for each destination suffices. This  allows the introduction of a technique to sequentially estimate the posterior distribution $p(T|D)$ of the arrival time $T$ from the available partial trajectory of the tracked object. Within the formulation adopted here, possible destinations may  also be  specified as  (Gaussian) distributions, with a mean and covariance,  in order to model endpoints with a non-zero spatial extent; point destinations can be modelled by setting the covariance to zero. Finally,  we  conduct simulations to illustrate the
inference capability of the bridging-distribution-based predictors  using real pointing data (HCI) and synthetic vessel tracks (MSA).
\subsection{Related Work}
One of the first techniques to incorporate  predictive information on the target endpoint to improve the accuracy of the tracking results  was proposed in \cite{castanon1985algorithms} and motivated various subsequent studies such as \cite{baccarelli1998recursive}. It assumes prior knowledge of the time of arrival at destination to devise a destination-aware tracker.  In this paper, the objective is to predict the intended destination of the tracked object, using motion models that are also inherently dependent on the endpoint.  This can be inferred using   a multiple Kalman-filter-based solution, even  without  imposing knowledge of $T$. It is noted that unlike the widely used interacting multiple models (IMM)  \cite{li2005survey,blom1988interacting}  and generalised pseudo-Bayesian \cite{tugnait1982adaptive} approaches for manoeuvering targets,  here we construct a bridge model per nominal destination and no interaction or switching among models is  applied. This is based on the premise that the tracked object intent is set well in advance of reaching its endpoint,  leading to  simple and low complexity algorithms. 
   
More recently,  destination-aware trackers that facilitate  inference of the object state $X_{t}$ followed by an additional mechanism to determine its endpoint are presented in  \cite{wang2011intent,fanaswala2013detection,nunez2013hard}, succeeding the work in \cite{bobick1998action}. The object trajectory is modelled as  a discrete  stochastic reciprocal  or  context-free grammar   process, which can be regarded as  non-causal generalisations of Markov processes. The state space is discretised within predefined  regions, for example, the spatial dimensions are divided into finite grids. The target can accordingly pass through a finite number of zones. Here, in contrast,  we adopt  continuous state space  models with bridging distributions that do not impose any discretisation/restrictions  on the path the tracked object has to follow to reach its endpoint. This formulation is particularly important in applications where discretisation of the spatial area is burdensome, e.g. tracking objects in 3D as with free hand pointing gestures or surveying a large geographical  area for MSA. The method proposed here  provides a simple,  effective solution to the intent prediction problem compared with those in \cite{fanaswala2013detection,nunez2013hard}; it also combines the destination prediction and tracking operations.  

Intent inference is often treated within  the context of anomaly detection, e.g. \cite{pallotta2013vessel,nunez2013hard,piciarelli2008trajectory,ristic2008statistical,fanaswala2013detection,derpanis2013action,pallotta2014context}. A\ common approach is to  categorise each trajectory of the tracked object(s)  as either normal or anomalous, i.e. classification techniques.
 This follows defining (or learning from recorded data) a \textit{pattern of life} that constitutes ordinary behaviour. Deviations from this are considered to indicate an anomaly, assuming adequate data association algorithms \cite{ristic2008statistical,pallotta2013vessel}. For example, a support vector machine (SVM) based method is introduced in [5] to classify deviant trajectories. A model-based technique, based on a tuned hidden Markov model, is utilised in \cite{pallotta2014context} to characterise a  \textit{pattern of life} and predict the future position of a conforming target.  Similarly, in \cite{fanaswala2013detection} and related work,  `tracklets', which are  sub-patterns comprising
a trajectory, are defined as  a means to capture a semantic
interpretation of complex patterns such as anomaly or intent. In this paper,  a probabilistic model-based formulation is proposed to tackle the   intent inference problem rather than anomaly detection, with the posterior probabilities of various intents sequentially  calculated using   bridging distributions. 

The benefits of predicting the intended item on a Graphical User Interface (GUI) early in a pointing task  are widely recognised in HCI, e.g. \cite{mcguffin2005fitts,asano05predict,lank2007endpoint,ziebart12}. Most existing  algorithms focus on pointing via a mechanical device, such as a mouse, in a 2D set-up.  In \cite{ahmadCyberTrans2015}, 2D--based predictors are shown to be unsuitable  for pointing tasks in 3D. For instance, the linear-regression methods  in \cite{asano05predict} and \cite{lank2007endpoint} assume that the the  destination is always  located along the path followed by the pointing object, which is rarely true in free hand pointing gestures.   Endpoint inference based on modelling the pointing  movements as a linear destination reverting  process  is considered in \cite{ahmadCyberTrans2015}. Compared to \cite{ahmadCyberTrans2015}, the  bridging-based-solution developed here is more general and more robust to variability in the target behaviour, leading to superior prediction results. 

Finally, data driven prediction/classification techniques, such as in  \cite{ziebart2009planning, ziebart12,bando2013unsupervised,kitani2012activity}, rely on a dynamical  model learnt from recorded tracks/behaviour. This often entails a high computational cost and necessitates substantial parameters training from  complete data sets (not always available). In this paper, as is common in tracking applications \cite{li2003survey,li2005survey,BarShalomBook2011}, known dynamical and sensor models (e.g. due to practical physical constraints), albeit with unknown parameters,  are presumed, i.e. a state-space-modelling approach. It requires minimal training, is amenable to parallelisation and is computationally efficient, yet  delivers a competitive performance.

\subsection{Paper Outline}
The remainder of this paper is organised as follows. In Section \ref{sec:ProblemStatement}, we formulate the tackled problem and  objectives. A range of possible motion models are outlined  in Section \ref{sec:models} and the prior of the state value at destination, i.e. $X_{T}$, is addressed. In Section \ref{sec:inference}, the bridging-based-prediction is introduced  and pseudo-code for the proposed algorithms is provided. The performance of the proposed techniques is evaluated in Section \ref{sec:Results} using real   free hand pointing gesture data and synthetic vessel tracks. Finally, conclusions are drawn in Section \ref{sec:Conclusions}. \section{Problem Statement}
\label{sec:ProblemStatement} 
Let $\mathbb{D}=\left\{ d : d=1,...,N\right\}$ be the set of $N$ nominal destinations, e.g. harbours where a vessel can dock or selectable \  icons displayed   on a touchscreen. The time instant  the tracked object reaches the \textit{a priori} unknown intended destination $D$ is denoted by  $T$. Whilst no  assumptions are made about the layout of   $\mathbb{D}$,  each endpoint is modelled as an extended region, e.g. GUI icons or harbour, rather than a single point as in \cite{ahmadCyberTrans2015}. Hence, the $d^{th}$ destination is defined by the (Gaussian)  distribution  $d\backsim\normdist{a_d}{\Sigma_d}$; see Section \ref{sec:bridging}. 

The objective here is to dynamically determine the probability of each possible endpoint being the intended destination,  
\begin{equation}\label{eq:Likelihoods}
\mathcal{P}(t_{n})=\left\{p(D=d\mid y_{1:n}): d=1,2,...,N\right\}
\end{equation}
 where $y_{1:n}\triangleq \left\{ y_{1},y_{2},...,y_{n}\right\}$  is the available partial trajectory or observations of the tracked object at the time instant $t_{n}\leqslant T$. For example, as in \cite{ahmadCyberTrans2015}, we have  $y_{k}=\left[ \hat x_{t_k}~~\hat y_{t_k}~~\hat z_{t_k}\right]'$ is the Cartesian  coordinates of the pointing finger at $t_{k}$ as captured by the pointing-gesture tracking device. Each observation is assumed to be  derived from a true, but unknown, underlying target state   $X_{t_n}$  at time $t_n$, which can include position, velocity and higher order kinematics. Given $\mathcal{P}(t_{n})$,  the Maximum \textit{a Posteriori} (MAP)\ estimate     
\begin{equation}\label{eq:MAP}
 \hat D(t_{n})=\underset{d=1,2,...,N}{\arg\max} \;\; p(D=d\mid y_{1:n}),
\end{equation}    
is an intuitive approach to determine the intended endpoint $D$ following the   arrival of the new observation $y_{n}$.  It should be noted that other decision criteria can be adopted, although this is not treated here. Therefore, the proposed probabilistic endpoint  prediction relies on a  belief-based inference, i.e. calculating $\mathcal{P}(t_{n})$ in (\ref{eq:Likelihoods}), followed by a classifier, e.g.  (\ref{eq:MAP}). 

As well as establishing  $D$ at time $t_{n}$, a refined estimate of the system latent state $X_{t_{n}}$ and   its predicted future values, i.e. $X_{t^*}$ for $t^*>t_n$, and time of arrival at $D$ are sought. A successful prediction at $t_{n}$ of the tracked object's final destination and/or its future trajectory can reveal its intention, alerting an operator  $T-t_{n}$  in advance of any potential conflict or reducing the pointing time by $T-t_{n}$ in an HCI context.      
\section{System Models}
\label{sec:models}

A linear and Gaussian motion model for the evolution of the physical state $X_t$ of the tracked object is assumed throughout this paper.  Whilst the system governing the target dynamics does not change over time, it does depend on the object eventually reaching its destination $D\in\mathbb{D}$. Conditioned on knowing the endpoint $D=d$, this leads to a linear time-invariant Gaussian system such that  the relationship between the system state at times $t$ and $t+h$ can be written as
\begin{equation}\label{eq:model}
X_{t+h} = F(h,d)X_t + M(h,d) + \varepsilon_t
\end{equation}
with $\varepsilon_t\sim\normdist{0}{Q(h,d)}$. The matrices $F$ and $Q$ as well as the vector $M$, which together  define the state transition from one time to another, are functions of the time step $h$ and the destination $d$. 

The  $n^{th}$ observation $y_n$  is modelled as a linear function of the time $t_n$ state perturbed by additive Gaussian noise, 
\begin{equation}\label{eq:obs}
y_n = GX_{t_n} + \nu_n
\end{equation}
where $\nu_n \sim \normdist{0}{V_n}$.  No assumption is made about the observation arrival times $t_n$ and irregularly spaces, asynchronous observations can naturally be incorporated within the framework presented here. Based on (\ref{eq:model}) and (\ref{eq:obs}), a Kalman filter can be utilised to calculate both the posterior distribution of the  latent state and the observation likelihood for the current set of measurements  $y_{1:n}$ \cite{haug2012}, conditioned on knowing $d$. The computationally efficient Kalman filter  is particularly desirable since   running, concurrently, multiple Kalman filters for $d\in\mathbb{D}$,  in real-time, is plausible, even in  settings where limited computing power is available, such as on a vehicle touchscreen or a portable battery-powered system.     

To condition on the destination, the  inference framework  requires that the transition density of such models can be calculated both from  one observation to the next (i.e. from time instant $t_{n-1}$ to $t_{n}$) and from the current observation time to the arrival time at the intended endpoint (i.e. from $t_{n}$ to  $T$).  Continuous-time motion models are therefore a natural choice, where the tracked object dynamics is represented by a   continuous-time stochastic differential equation (SDE). This SDE can be integrated to obtain a transition density over any  time interval.  For Gaussian linear time invariant (LTI)  models, this integration is analytically tractable, giving   transition functions of the form in (\ref{eq:model}).  This class of models  includes many  widely used ones, e.g.    (near) constant velocity (CV) and   (near) constant acceleration (CA) models, as well as   the linear destination reverting models given below.

The early work in \cite{ahmadCyberTrans2015}  employs LDR models in a forward sense, without using the destination state  as conditioning information. In this paper, by contrast,  a prior probability distribution on the object state  at its destination $X_{T}$ is used directly as conditioning information in the inference procedure. This is achieved by using  bridging distributions to introduce the longer term dependencies in the target trajectory (see Section \ref{sec:bridging}).  A similar idea is explored in the preliminary study in \cite{ahmad2015ICASSP} using the CV model, in which  the endpoint information does not feature in the motion model. That work also specifies each destination as a single point (rather than a prior distribution)  and    applies a two step modified  Kalman filter  to infer $D$, which is more computationally demanding and obscure compared with the solution presented here. 

\begin{figure}[t]
\centering
\includegraphics[width=1\linewidth]{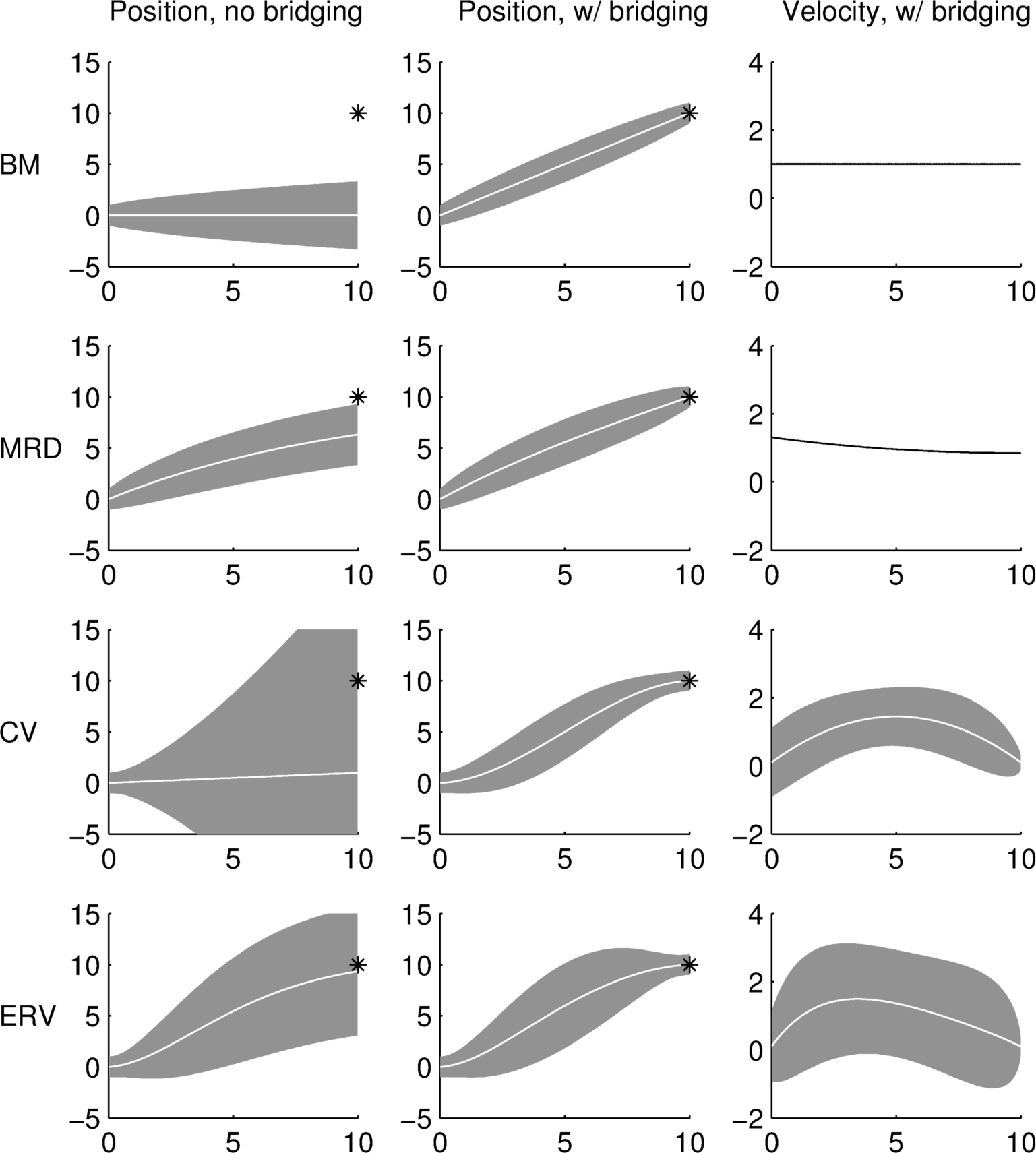}
\caption{\small Predictive distribution at $t=0$ of position without bridging (first column), predictive position with bridging (second column), and predictive  velocity with bridging (third column) for the BM, MRD, CV and ERV models in 1D. The $x$-axis shows time $t$ and $y$-axis depicts position/velocity.  The destination (black star), is located at position 10 at time $T=10$.  Grey shading shows one standard deviation, white line depicts the mean (except for BM and MRD velocity, for which the black line shows the implied mean of velocity only).  Parameters for the models are $\sigma = 1$ (all models), $\lambda=0.3$ (MRD), $\eta=0.1, \rho=0.5$ (ERV). }
\label{fig:distribs}
\end{figure}

In the remainder of this section, we describe the LDR  models of \cite{ahmadCyberTrans2015} for completion, other  related motion processes and the prior of $X_{T}$. The predictive position and velocity distributions of  selected motion models, with and without the use of bridging (i.e. conditioning on $X_{T}$) are depicted in Figure \ref{fig:distribs} in the one dimensional case.  The figure demonstrates the substantial effect of introducing the bridging assumptions on the prediction results of various motion models. It clearly shows how these distributions more accurately model the predicted state at the endpoint.   It is worth noting that any continuous-time LTI model  could be used directly within the proposed inference framework, as long as the system can be expressed by \equations{model}{obs}.   
\subsection{Mean Reverting Diffusion (MRD)}\
\label{sec:MRD}
The position of the tracked object for the Mean Reverting Diffusion (MRD) model follows  an Ornstein-Uhlenbeck process \cite{gillespie1996exact}, with its mean being the  destination.     The intuition behind MRD is that a target heading towards a particular endpoint will revert towards it. The reversion strength is stronger the further the target is from its destination.  This produces the  predictive position and velocity dynamics of the form displayed in the second row of Figure \ref{fig:distribs}. 

The   state $X_t$ of the mean reverting diffusion  model consists of only the target position in each of the $s$ spatial dimensions. For  endpoint $d$ (located at position   $p_d$),  the SDE is  given by
\begin{equation}\label{eq:MRD_SDE}
dX_{t}=\Lambda\left( p_{d}-X_{t}  \right)dt +\sigma d w_{t},
\end{equation}
where $\Lambda$  is a diagonal matrix with diagonal elements $\{\lambda_i\}_{i=1}^{s}$ that set the reversion coefficient in each  dimension. The diagonal matrix  $\sigma$  specifies the standard deviation of the dynamic  noise  and  $w_t$ is a  standard (unit variance) $s$-dimensional Weiner process.  The diagonal structure of $ \sigma$ implies that the noise is independent in each spatial  dimension, which is a common assumption   in  tracking  \cite{li2003survey}. However, this can be relaxed as shown  in \equatn{MRD-non-diagonal} below.

As in \cite{ahmadCyberTrans2015}, by integrating (\ref{eq:MRD_SDE})  from $t$ to $t+h$, we obtain a state transition function in the form of equation (\ref{eq:model}) with
\eqn{MRD}
{
F_\text{MRD}(h,d) &= e^{-\Lambda h},\nonumber\\
M_\text{MRD}(h,d) &= (I_s-e^{-\Lambda h})p_d,\\
Q_\text{MRD}(h,d) &= \frac{1}{2}\left[I_{s}-e^{-2\Lambda h} \right]\Lambda^{-1} \sigma^{2}\nonumber,
}
where $I_s$ is the $s\times s$ identity matrix.  For non-diagonal $\sigma$, the $(i,j)$\thh element of $Q_\text{MRD}$ is given by
\begin{equation}\label{eq:MRD-non-diagonal}
Q_\text{MRD,ij} = \frac{(\boldsymbol\sigma\boldsymbol\sigma')_{ij}}{\Lambda_{ii}+\Lambda_{jj}}\left[1-e^{-(\Lambda_{ii}+\Lambda_{jj})h}\right],
\end{equation}
where $z'$ is the transpose of the vector/matrix $z$. A special case of the MRD model occurs when $\Lambda=0_{s}$ (with $0_s$ being the $s\times s$ matrix of zeros), in which case  the dynamics of the target position follow a Brownian motion (BM).  In this case, the  $F$, $M$ and $Q$ matrices in (\ref{eq:model}) become
\eqn{}
{
F_\text{BM}(h,d) &= hI_s,\\
M_\text{BM}(h,d) &= 0_{s\times1},\\
Q_\text{BM}(h,d) &= h \sigma^{2},
}
where $0_{s\times1}$ is an $s \times 1$ column vector of zeros.

If measurements $y_n$ are direct noisy observations of the tracked object position, the  observation matrix $G$ in \equatn{obs} is simply the identity matrix,  $G = I_s$.

\subsection{Equilibrium Reverting Velocity (ERV)}
\label{sec:ERV}

In the ERV model, introduced in \cite{ahmadCyberTrans2015} for tracking and destination inference for pointing tasks on in-car displays,    the system state  $X_t$ at time $t$ is  a $2s\times1$ vector, arranged as $[x_1,...,x_s, \dot x_1,...,\dot x_s ]'$, where $x_i$ is the position in spatial dimension $i$, and $\dot x_i$ is the velocity in that  dimension. 
Under this model, the SDE governing the evolution of the tracked object state
is
\begin{equation}\label{eq:ERV_SDE}
dX_t=A\left(\mu_{d}-X_t \right)dt+\sigma  d w_{t}, 
\end{equation}
where the mean $ \mu_{d} = [p_d ',0'_{s \times 1}]'$  contains the position $p_d$ of destination $d$;    $w_{t}$ is a standard $s$ dimensional Wiener process and $B=[0_s, \sigma]$ is a $2s\times s$ matrix that controls the noise which affects the velocity components of the process (modelling random forces acting on the target).  In this case, $\sigma$ can be given by the $s\times s$ Cholesky decomposition of the velocity noise covariance  $\Sigma$, such that $\Sigma=\sigma\sigma'$. As with the MRD model, a common choice is a diagonal  $\sigma$, i.e. independent noise in each spatial dimension.  
The matrix $A$ is given by
\eqn{}
{
A = \inmat{0_s & -I_s\\ \eta & \rho},
}
where $\rho$ is an $s\times s$ diagonal matrix of the drag coefficients (can be assumed to be the same across all dimensions) and $\eta$ is a $s\times s$ diagonal matrix of the mean reversion strengths in each spatial dimension.

A physical interpretation of the ERV model is   that the destination $d$ exerts an attractive force on the tracked object, of strength proportional to the distance separating them.  This can be viewed  as  a linear spring of zero natural length connecting the target to its endpoint. A drag term proportional to the velocity of the target is also included, allowing the velocity profile of the tracked object (e.g. pointing-finger-tip in a free hand pointing gesture) to be correctly modelled; see Figure \ref{fig:distribs} (fourth row) and \cite{ahmadCyberTrans2015}.  

Integrating the SDE in (\ref{eq:ERV_SDE}) from $t$ to $t+h$ allows the system evolution to be expressed in the form in \equatn{model} with
\eqn{} 
{
F_\text{ERV}(h,d) &= e^{-Ah},\\
M_\text{ERV}(h,d) &= (I_{2s}-e^{-A h})\mu_d,\\
Q_\text{ERV}(h,d) &= \int_{t}^{t+h}e^{-A(t+h-v)} \sigma \sigma 'e^{-A'(t+h-v)}dv.
} 
The covariance matrix $Q_\text{ERV}(h,d)$ can be calculated by Matrix Fraction Decomposition \cite{sarkka2006recursive}  where
\eqn{}
{
Q_{\text{ERV}}(h,d) = JK^{-1},
}
\eqn{}
{
\inmat{J\\K} = \exp\brackets{\inmat{-A & \sigma \sigma '\\0_{2s} &  A'}h}\inmat{0_{2s} \\ I_{2s}}.
}

A special case of the ERV model occurs when $\eta$ and  $\rho$ are both zero.  In this scenario, the model reduces to the widely used (near) constant velocity model with 
\eqn{}
{
F_\text{CV}(h,d) &= \inmat{I_s & hI_s \\ 0_s & I_s},\\
M_\text{CV}(h,d) &= 0_{2s\times1},\\
Q_\text{CV}(h,d) &= \inmat{\sigma\sigma' \frac{h^3}{3} & \sigma\sigma' \frac{h^2}{2} \\\sigma\sigma' \frac{h^2}{2} & \sigma\sigma' h}.
}
If observations $y_n$ are direct noisy measurements of the target position, the observation matrix $G$ in \equatn{obs} is given by $G = \inmat{I_s & 0_s}$.
\subsection{State Value Prior at Destination}
\label{sec:bridging}

Knowing the intended destination $D$ of a tracked object gives information about the system state at some future time $T$.   This can be modelled by a prior probability distribution for   $X_T$ corresponding to the geometry of the destination,  since most endpoints are  extended regions (e.g. GUI\ buttons or harbours), rather than single points. In order to maintain the linear Gaussian structure of the  system, the following Gaussian prior on the  object state upon arrival at  destination at time $T$: 
\eqn{dest-prior}
{
p(X_T\mid D=d) = \normpdf{X_T}{a_d}{\Sigma_d},
}
is assumed. This effectively  models the destination as ellipsoidal, since the iso-probability surfaces of the Gaussian distribution are elliptical. The mean vector $a_d$ specifies the location/centre of the destination, for example, $a_d=p_d$ for the MRD model and $a_d= \mu_d$ from \equatn{ERV_SDE} for the ERV model. Whereas,  $\Sigma_d$ is a covariance matrix of the appropriate dimension, which sets the extent and orientation of the endpoint.  In the case of the ERV model, defining  the destination  also entails  specifying a distribution of the tracked object velocity at the destination. If this is unknown, a large prior variance can be used to model this uncertainty.
\section{Intent Inference Using Bridging Distributions}
\label{sec:inference}
\begin{figure}[b]
\centering
\includegraphics[width=0.9\linewidth]{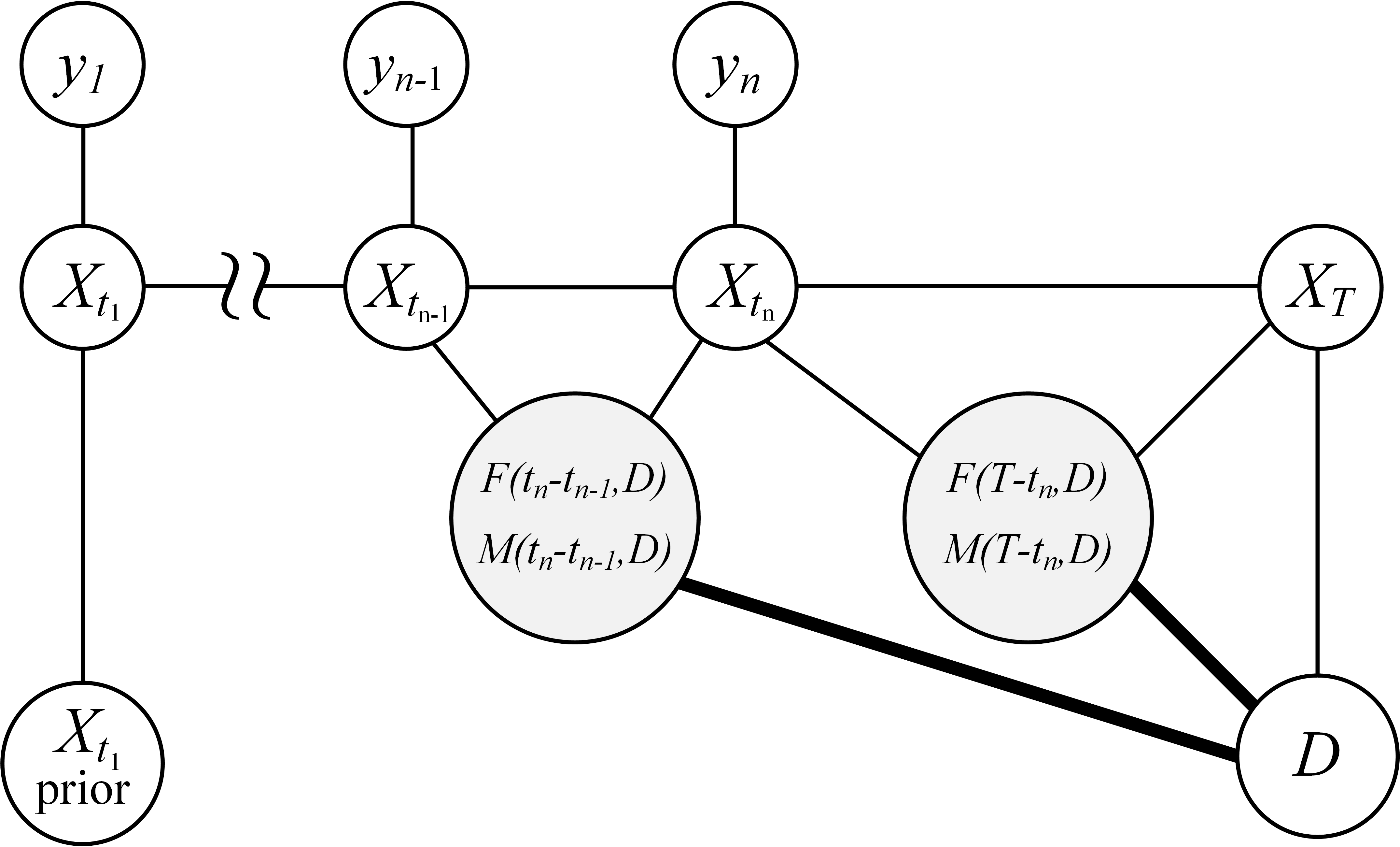}
\caption{\small The graphical structure of the system after $n$ observations.  The destination $D$  plays a similar role to the prior distribution of $X_{t_1}$ in addition to affecting the state transition function.  Heavy lines indicate a deterministic relationship (early state transition matrices  are not shown in the figure).}
\label{fig:structure}
\end{figure}

For motion models of the form in \equatn{model},  and conditioning on  a given destination $d\in \mathbb{D}$ as well as  arrival time $T$, the posterior of the system state can be expressed by   $p(X_{t_n}\mid y_{1:n}, T, D=d)$ and  the observation likelihood is   $p(y_{1:n}\mid T, D=d)$ after $n$ measurements. The graphical structure of this system is depicted in Figure \ref{fig:structure}.  The intended  destination $D$ influences the state at all times via the  reversion built into the LDR\ motion models (for  non-destination reverting models, e.g.  CV\ and BM, the endpoint only affects the final state).  Nonetheless, the inclusion of the prior on $X_{T}$ in \equatn{dest-prior} changes the system dynamics compared to those of the MRD and ERV models in \cite{ahmadCyberTrans2015}.  Under the assumption that the destination is known, the posterior distribution of the state changes from a random walk to a bridging distribution. This is clearly visible in Figure 2, especially in terms of producing  consistent and meaningful  predictions.   
\subsection{Filtering and Likelihood Calculation}\label{sec:filtering}
An elegant way to  filter for $X_{t}$  with a destination prior  is to extend the latent system state to incorporate $X_{T}$, thus, it becomes $Z_t =[X_t' ~X_T']'$. Filtering is then carried out for $p(Z_{t_n}\mid y_{1:n},D=d, T)$, permitting the calculation of the observation likelihood $p(y_{1:n}\mid D=d, T)$. The key to developing this filter is to calculate the transition density $p(Z_{t+h}\mid Z_t, D=d,T)$, given by
\begin{align}
p(Z_{t+h}\mid Z_t,D=d,T)&=p(X_T\mid X_{t+h},X_T,D=d,T) \nonumber
\\&~~~~~~~\times p(X_{t+h}\mid X_t,X_T,D=d,T)\nonumber\\
&=p(X_{t+h}\mid X_t,X_T,D=d,T).
\end{align}
This follows because  $p(X_T\mid X_{t+h},X_T,D=d,T)=1$ for any value of $X_T$, due to the fact that the state $X_T$ is included in the conditioning information via $Z_t$. The distribution $p(X_{t+h}\mid X_t,X_T,D=d,T)$ is given by
\begin{align}\label{eq:tx-density1}
&p(X_{t+h}\mid X_t,X_T,D=d,T) \propto p(X_T\mid X_{t+h},D=d,T) \nonumber \\
& ~~~~~~~~~~~~~~~~~~~~~~~~~~~~~~~~~~~~~~~~\times p(X_{t+h}\mid X_t,D=d,T)\nonumber\\
&=\normpdf{X_{T}}{F_xX_{t+h}+M_x}{Q_x}\
\normpdf{X_{t+h}}{F_hX_t+M_h}{Q_h}\nonumber \\ 
\end{align}
where the $F$, $M$ and $Q$ matrices are taken from the motion models in \ssection{models}, with
\eqn{}
{
F_x &= F(T-t-h,D)  &F_h &= F(h, D), \\
M_x &= M(T-t-h,D)  &M_h &= M(h, D), \\
Q_x &= Q(T-t-h,D)  &Q_h &= Q(h, D).
}
This comes from the system structure  in \fig{structure}, and the fact that for the continuous-time integrable models used here, the state transition density can be calculated over any time period.  

The following Gaussian identity simplifies the state transition density in \equatn{tx-density1}:\eqn{gaussian-id}
{
\normpdf{x}{\mu_1}{\Sigma_1}\normpdf{\mu_2}{Lx}{\Sigma_2} = z\normpdf{x}{\mu_*}{\Sigma_*}
}
where $L$ is a matrix of appropriate size, $z$ is a normalizing constant that does not depend on $x$, and
\eqn{}
{
\Sigma_* &= \brackets{\Sigma_1^{-1}+L'\Sigma_2^{-1}L}^{-1},\\
\mu_*    &= \Sigma\brackets{\Sigma^{-1}_1\mu_1 +L'\Sigma_2^{-1}\mu_2}.
}
This leads to
\eqn{}
{
p(X_{t+h}\mid X_t,X_T,T,D=d) = \normpdf{X_{t+h}}{c_t}{C_t},
}
where
\eqn{}
{
C_t &= (Q_h^{-1} + F_x'Q_x^{-1}F_x)^{-1},\\
c_t &= C_t \left[Q_h^{-1}(F_h X_t + M_h) + F_x'Q_x^{-1}(X_T-M_x)\right],\\
  &= H _tZ_t +m_t,
}
with, for an $r$-dimensional state vector $X_t$, $H_t$ a $r\times2r$ matrix and $m_t$ a $r\times1$ vector of the form
\eqn{}
{
H_t &= [C_tQ_h^{-1}F_h, ~C_tF_x'Q_x^{-1}],\\
m_t &= C_t(Q_h^{-1}M_h-F_x'Q_x^{-1}M_x).
}
This allows the state transition with respect to $Z_t$ to be written as
a linear Gaussian transition of the form
\eqn{Zstatex}
{
Z_{t+h}&=R_tZ_t+\tilde{m}_t +\gamma_t\label{eq:Zstate}\\
\gamma_t &\sim \normdist{0}{U_t}.\nonumber
}
where 
\begin{equation}\label{eq:RUm}
R_t = \inmat{H_t \\ P_T}, \quad\tilde{m}_t = \inmat{m_t\\0_r},  \quad U_t = \inmat{C_t & 0_r \\ 0_r & 0_r},
\end{equation}
and $P_{T}=\inmat{0_r & I_r}$.  With respect to this extended system, the $k$-dimensional observation vector at each observation time is given by
\eqn{Zobs}
{
y_n &= \tilde G Z_{t_n}+ \nu_n
}
with $\tilde G = [G, ~0_{k\times r}]$ and where $G$ and $\nu_n$ are as those in \equatn{obs}.
\begin{algorithm}
\begin{algorithmic}
\STATE \textbf{Notation:} $\{\ell_n, \hat Z_n, \Sigma_n  \}=\text{KF}(y_n, \hat Z_{n-1}, \Sigma_{n-1},R_t,U_t,\tilde G)$ 
\STATE \textbf{Input}: Observation $y_n$; previous posterior state mean estimate$^\dagger$ $\hat Z_{n-1}$; previous posterior state covariance$^\dagger$ $\Sigma_{n-1}$; state transition matrix $R_t$ and covariance $U_t$ from \equatn{Zstate}; observation matrix $\tilde G$ from \equatn{Zobs} 

\STATE \textbf{Predict$^\dagger$:}
\STATE ~~ $\hat Z_{n|n-1} = R_t \hat Z_{n-1}+\tilde{m}_t$
\STATE ~~ $\Sigma_{n|n-1} = R_t\Sigma_{n-1}R'_t + U_t$
\STATE \textbf{PED  Calculation:}
\STATE ~~$~\ell_n =\normpdf{y_n}{\tilde G\hat Z_{n|n-1} }{\tilde G \Sigma_{n|n-1} \tilde G'}$

\STATE \textbf{Correct:}
\STATE ~~ $K = \Sigma_{n|n-1}\tilde G'(\tilde G \Sigma_{n|n-1} \tilde G'+V_n)$
\STATE ~~ $\hat Z_n = \hat Z_{n|n-1} + K(y_n-\tilde G \hat Z_{n|n-1}$
\STATE ~~ $\Sigma_n = (I_r-K\tilde G)\Sigma_{n|n-1}$

\STATE \textbf{Output:} PED $\ell_n = p(y_n\mid y_{1:n-1},D=d,T)$; posterior mean of state at time $t_n$, $\hat Z_n$; posterior state covariance $\Sigma_{n-1}$.

\STATE $^\dagger$For first observation $y_1$, skip predict step and use prior mean and covariance for  $Z_{t_1}$ from \equatn{Zprior} in place of $\hat Z_{n|n-1}$ and $\Sigma_{n|n-1}$ in likelihood calculation and correction steps.

\end{algorithmic}
\caption{Conditioned Prediction Error Decomposition and State Inference - Kalman Filter (single iteration)}
\label{alg:1}
\end{algorithm}

\Equations{Zstate}{Zobs} form a standard linear Gaussian system, albeit  with a degenerate state transition covariance matrix. Thus, a standard Kalman filter can be applied  to calculate the conditioned posterior filtering distribution of the state and the conditioned prediction error decomposition (PED), $p(y_n\mid y_{1:n-1},D=d,T)$.   This latter is shown in Sections \ref{sec:Destination-Inference} to \ref{sec:state-prediction} to be key to destination inference, because it allows the destination and end-time conditioned observation likelihood of the partially observed track to be calculated recursively as $p(y_{1:n}\mid D=d,T)=p(y_n\mid y_{1:n-1},D=d,T)p(y_{1:n-1}\mid D=d,T)$. The algorithm for a single iteration of the Kalman filter is given in  \Algo{1}. It requires the prior $Z_{t_1} = [X_{t_1}', X_T']'$ at the first observation time. The prior on $X_{t_1}$ is the standard prior on the initial state.  The prior on $X_T$ is derived from the destination as described in \ssection{bridging}, and can be neatly incorporated into the Kalman filter.  Both these priors must be Gaussian, and in most modelling scenarios they will be independent of each other, giving a prior for $Z_{t_1}$ of the form
\eqn{Zprior}
{
p(Z_{t_1}\mid T,D=d) = \normpdf{\inmat{X_{t_1}\\X_T}}{\inmat{\mu_1\\a_d}}{\inmat{\Sigma_1 & 0_r \\0_r & \Sigma_d}},
}
where $\mu_1$ and $\Sigma_1$ specify the initial prior on $X_{t_1}$, i.e. $p(X_{t_1}\mid D=d) = \normpdf{X_{t_1}}{\mu_1}{\Sigma_1}$;~$a_d$ and $\Sigma_d$ come from the destination prior in \equatn{dest-prior}.

\subsection{Unknown Arrival Times}
So far, it has been assumed that the arrival time at the destination $T$ is known \emph{a priori}.  Often, this is not a realistic assumption and we are interested in the probability of the
tracked object arriving at a destination at \textit{any} time within some time interval.
In this case, the unknown arrival time  $T$ is treated as a random variable, which must be integrated over  in order to infer the destination of the tracked object.
 
The observation likelihood with unknown arrival time is given by $p(y_{1:n}\mid D) $.   This can be calculated by integrating the arrival-time-conditioned likelihood calculated by the Kalman filter in the preceding  section (see \Algo{1}) over all possible arrival times according to
\begin{equation}\label{eq:integral}
p(y_{1:n}\mid D) = \int_{T\in\mathcal{T}} p(y_{1:n}\mid T,D)p(T\mid D) dT,
\end{equation}
where $p(T\mid D)$ is the a prior distribution of arrival times for destination $D$ and $\mathcal{T}$ is the time interval of possible arrival times $T$ ($D=d$ is here replaced by $D$ for notational brevity).  For example, arrivals might be expected uniformly within some time period $[t_a,t_b]$, giving  $p(T\mid D) = \mathcal{U}(t_a,t_b)$.

In most cases,  $p(y_{1:n}\mid T,D)$ after $n$ observations is a nontrivial function of  $T$ resulting in intractable integrals.  A numerical approximation to the integral in \equatn{integral} can be obtained via numerical quadrature, which is viable since the arrival time is a one-dimensional quantity.  The approximation requires  multiple evaluations of the arrival-time-conditioned-observation-likelihood (for various arrival times $T$) for each of the $N$  nominal endpoints. Therefore, in time-sensitive applications, it is likely that only  a few quadrature points $q$ can be  used. 

Whilst adaptive quadrature schemes might  seem appealing for their efficiency and accuracy, they are not easily applied to this problem. For a given  arrival time, only a single iteration of the Kalman filter as in \Algo{1} need be run (per destination $d\in \mathbb{D}$), following the arrival of a new observation.  Thus, if a fixed quadrature is utilised with $q$ quadrature points, $Nq$ Kalman filter iterations of the form in \Algo{1} are required per observation arrival.  For adaptive quadrature, it will, in general, be  necessary to calculate the observation likelihood for a different set of arrival times at each step.  This, however, requires the Kalman filter be re-run from scratch for all $n$ available observations, imposing $nNq$ iterations.  Hence, simple fixed grid quadrature schemes are employed here. A Kalman filter iteration as in \Algo{1}   is not computationally intensive  and each of the $Nq$ iterations, required per observation, can be performed in parallel. For arrival times before the current observation time $T_i<t_n$, the arrival-time-conditioned observation likelihood is taken to be zero, such that $p(y_{1:n}\mid T=T_i<t_n,D) = 0$.

Here, we apply a Simpson's rule quadrature scheme, with  an odd number $q$ of evenly spaced quadrature points, $T_1,...,T_q$.  This approximates the integral in \equatn{integral} by
\begin{align}\label{eq:simpson}
&p(y_{1:n}\mid D) \approx \frac{T_q-T_1}{3(q-1)}\bigg[p(y_{1:n}\mid T=T_1, D) \nonumber
\\&~~~~+ p(y_{1:n}\mid T=T_q, D) +4\sum_{i=1}^{(q-1)/2} p(y_{1:n}\mid T=T_{2i}, D) \nonumber
\\&~~~~~~~~~~~~~~~~~~+2\sum_{i=1}^{(q-1)/2-1} p(y_{1:n}\mid T=T_{2i+1}, D)\bigg].
\end{align}
Other numerical integration schemes such as Gaussian quadrature can  be employed. It may also be desirable to use unequal intervals between quadrature points, e.g. if  the  distribution of arrival times is heavily peaked in particular regions.

\subsection{Destination Inference}\label{sec:Destination-Inference}
The  posterior distribution of the nominal destinations, i.e. $p(D=d\mid y_{1:n}), d\in \mathbb{D}$ in \equatn{Likelihoods}, at the time instant $t_{n}$ can be expressed via Bayes' theorem by  
\eqn{D-posterior}
{
p(D=d\mid y_{1:n}) \propto p(y_{1:n}\mid D=d)p(D=d).
}
The discrete probability distribution $p(D=d)$ defines a prior over all possible destinations; it is independent of the current track  and can be obtained from contextual information, historical data, etc. Alternatively, a non-informative prior can be used where  $p(D=d)=1/N$ for all $d\in\mathbb{D}$. 

\Algo{2} shows how the posterior distribution over destinations $p(D=d\mid y_{1:n})$\ in (\ref{eq:Likelihoods})  can be inferred sequentially given a series of observations $y_{1:n}$.   It begins by initializing the running likelihood estimate $L_0^{(d)}$ for each  $d\in \mathbb{D}$ and the current  posterior state mean $\hat Z^{(d,i)}_0$ as well as covariance $\Sigma^{(d,i)}_0$ for each destination $d$ and quadrature point $i$ (corresponding to arrival time $T_i$) to their priors.   After each observation, the Kalman filter iteration in \Algo{1} is utilised to calculate the one step arrival-time-conditioned-observation PED $\ell_n^{(d,i)}=p(y_n\mid y_{1:n-1}, T=T_i,D=d)$ for each destination $d\in\mathbb{D}$ and quadrature point $i$. This is then used to calculate the overall arrival-time-conditioned observation likelihood
\begin{align}
&L_n^{(d,i)} = p(y_{1:n}\mid T=T_i, D=d), \nonumber \\
&= p(y_n\mid y_{1:n-1}, T=T_i, D=d)p(y_{1:n-1}\mid T=T_i, D=d),\nonumber\\
&= \ell_n^{(d,i)} \times L_{n-1}^{(d,i)}.
\end{align}
The Kalman iteration also determines the corresponding updated posterior state mean $\hat Z^{(d,i)}_n$ and covariance $\Sigma^{(d,i)}_n$, which are necessary for the next steps.
\begin{algorithm}
\begin{algorithmic}
\STATE \textbf{Input:} Observations $y_{1:N}$
\STATE \textbf{Initialize:} Set $L_0^{(d,i)}=1$ and set $\hat Z^{(d,i)}_0$, $\Sigma^{(d,i)}_0$ to priors from \equatn{Zprior} for all $d\in\mathbb{D}$, $i=1,...,q$
\FOR{observations $n=1,...,N$} 
\FOR{destination $d\in\mathbb{D}$}
\FOR{quadrature point $i\in 1,...,q$}
\STATE Calculate $R_t^{(d,i)}$, $U_t^{(d,i)}$ in \equatn{Zstate} for observation time $t_n$, destination $d$ and arrival time $T_i$
\STATE Run Kalman filter iteration:
\STATE $~~\{\ell_n^{(d,i)}, \hat Z_n^{(d,i)}, \Sigma_n^{(d,i)}\}=$ 
\STATE ~~~~~~~~~~~~~~~$\text{KF}(y_n, \hat Z_{n-1}^{(d,i)}, \Sigma^{(d,i)}_{n-1},R^{(d,i)}_t,U^{(d,i)}_t,\tilde G)$
\STATE \COMMENT{$\ell_n^{(d,i)}$ is the PED\  for a known arrival time, i.e. $\ell_n^{(d,i)}=p(y_n\mid y_{1:n-1},D=d, T=T_i)$}
\STATE Update likelihood:  $L_n^{(d,i)} = L_{n-1}^{(d,i)}\times \ell_n^{(d,i)}$
\ENDFOR
\STATE Calculate likelihood approximation:
\STATE ~~~~~~~~~$P^{(d)}_n =  \text{quad}(L_n^{(d,1)}, L_n^{(d,2)},..., L_n^{(d,q)})$
\STATE \COMMENT{$P^{(d)}_n \approx p(y_{1:n}\mid D=d)$; quad is quadrature function}
\ENDFOR
\FOR{destination $d\in\mathbb{D}$}
\STATE $u_{d} = \frac{p(D=d)P_n^{(d)}}{\sum_{d\in\mathbb{D}}p(D=d)L_n^{(d)}}$
\ENDFOR
\STATE \textbf{Destination posterior after $n$\thh observation:} 
\STATE ~~~$p(D=d\mid y_{1:n})\approx u_d$ 
\ENDFOR

\end{algorithmic}
\caption{Destination Inference}
\label{alg:2}
\end{algorithm}

After running a Kalman iteration for all quadrature points, the quadrature function (and arrival time prior) is used on the likelihood $L_n^{(d,i)}$ calculated at each of these points for a destination $d$ in order to approximate the integral in (\ref{eq:integral}).  This gives the  observation likelihood $p(y_{1:n}\mid D=d)$ for each destination $d\in\mathbb{D}$,
\begin{equation}
{
p(y_{1:n}\mid D=d) \approx \text{quad}(L_n^{(d,1)}, L_n^{(d,2)},..., L_n^{(d,q)}).
}
\end{equation}
For a finite set   $\mathbb{D}$, the probability of any given destination as in (\ref{eq:Likelihoods}) and (\ref{eq:MAP}) can be determined  by evaluating the expression in \equatn{D-posterior} for each $d\in \mathbb{D}$, followed by the  normalisation  
\begin{equation}\label{eq:DestinationInference}
{
p(D=d\mid y_{1:n}) = \frac{p(y_{1:n}\mid D=d)p(D=d)}{\sum_{j\in\mathbb{D}}p(y_{1:n}\mid D=j)p(D=j)},
}
\end{equation}
which ensures that  the total probability over all possible destinations sums to 1.
In \Algo{2}, this is approximated (due to numerical quadrature) by $u_{d}\approx p(D=d\mid y_{1:n})$. For a fixed set of quadrature points, this estimated posterior can be updated sequentially after each observation, making it tractable for a large numbers of possible destinations ($N$)\ and quadrature points ($q$).
\subsection{Arrival Time Inference}\label{sec:arrival-time}
In addition to inferring the intended destination $D$ of the tracked object, it is also possible to infer a posterior distribution of the time at which the target is expected to reach its intended (unknown) endpoint.  For a specific  destination $D=d$, this is given by
\begin{equation}\label{eq:T-posterior1}
p(T\mid D=d, y_{1:t}) \propto p(y_{1:t}\mid T, D=d)p(T\mid D=d).
\end{equation}
Since the quadrature procedure used in Section \ref{sec:filtering} requires  calculation of the arrival-time-conditioned-likelihood $p(y_{1:t}\mid T=T_i, D=d)$ for a number of quadrature points $T_i$, a discrete approximation of the overall posterior can be obtained  almost without additional calculations via
\eqn{T-posterior-approx1}
{
p(T\mid D=d, y_{1:t}) \approx \sum_{i=1}^q w_i\diracat{T_i},
}
where $\diracat{T_i}$ is a Dirac delta located at $T_i$.  This is a weighted distribution over the quadrature points with  \eqn{}
{
w_i = \frac{p(y_{1:t}\mid T=T_i, D=d)p(T_i\mid D=d)}{\sum_{i=1}^qp(y_{1:t}\mid T=T_i, D=d)p(T_i\mid D=d)}.
}
These weights are  normalized evaluations of the expression in \equatn{T-posterior1}, calculated at each  $T_i$.  Normalization ensures that the approximate posterior distribution in (\ref{eq:T-posterior-approx1}) is a valid probability distribution that integrates to 1.

The posterior distribution of the arrival time  at \emph{any} destination can be calculated without significant further calculations by integrating over all endpoints (since $\mathbb{D}$ is a discrete set) as \begin{align}\label{eq:TimeOnAllDest}
&p(T\mid y_{1:t}) = \sum_{d\in\mathbb{D}} p(T,D=d\mid y_{1:t}) \nonumber \\
&\propto  \sum_{d\in\mathbb{D}} p(y_{1:t}\mid T, D=d)p(T\mid D=d) p(D=d).
\end{align}
Since the conditioned likelihood $p(y_{1:t}\mid T, D=d)$ is determined for each $d\in \mathbb{D}$ and  quadrature point $T=T_i$, $i=1,2,...,q$, the distribution in (\ref{eq:TimeOnAllDest}) can  be approximated by \eqn{}
{
p(T\mid y_{1:t}) \approx \sum_{i=1}^q v_i\diracat{T_i},
}
with the weights defined  by
\eqn{}
{
v_i = \frac{\sum_{d\in\mathbb{D}} p(y_{1:t}\mid T_i, D=d)p(T_i\mid D=d) p(D=d)}{\sum_{i=1}^q\sum_{d\in\mathbb{D}} p(y_{1:t}\mid T_i, D=d)p(T_i\mid D=d) p(D=d)}.
}
This relies on the assumption that the set of arrival time quadrature points are the same for each destination $d\in \mathbb{D}$.
\subsection{State Inference and Trajectory Prediction}
\label{sec:state-prediction}
Estimating the current state of the tracked object (e.g. position) and  predicting its future state can also be readily preformed within the framework presented. 
The posterior estimate of the target state $p(X_{t_n}\mid y_{1:t})$ is given by integrating over all possible destinations and arrival times, i.e.
\begin{align}\label{eq:Xcur}
p(X_{t_n}\mid y_{1:t}) &= \int\bigg[\sum_{d\in\mathbb{D}} p(X_{t_n}\mid y_{1:t},T,D=d)\nonumber
\\ & ~~\qquad \times p(T\mid D=d)p(D=d)\bigg]  dT.
\end{align}
The estimated   state at the time  instant $t_n$ conditioned on the arrival time $T_i$ and endpoint $d$ is $p(X_{t_n}\mid y_{1:t},T=T_i,D=d)$. It has a Gaussian distribution with mean and variance  given by $P_t\hat Z_n^\inb{d,i}$ and $P_t\Sigma_n^\inb{d,i}P_t'$, respectively, where, $P_t= \inmat{I_r & 0_r}$, i.e. the components of $\hat Z_n^\inb{d,i}$ and $\Sigma_n^\inb{d,i}$ in \Algo{2} corresponding to $X_{t_n}$ rather than $X_{T_i}$.

The distribution in (\ref{eq:Xcur})  can be approximated from the calculated arrival time and destination conditioned state estimates by the Gaussian mixture 
\eqn{}
{
p(X_{t_n}\mid y_{1:t}) &\approx \sum_{i=1}^q\sum_{d\in\mathbb{D}} u_{i,d}\normpdf{X_{t_n}}{P_t\hat Z_n^\inb{d,i}}{P_t \Sigma_n^\inb{d,i}P'_t}
}
where  weights in the above summation are given by \begin{equation}\label{eq:StateEstimate}
{
u_{i,d} = \frac{p(y_{1:t}\mid T_i, D=d)p(T_i\mid D=d) p(D=d)}{\sum_{i=1}^q\sum_{d\in\mathbb{D}} p(y_{1:t}\mid T_i, D=d)p(T_i\mid D=d) p(D=d)}.
}\end{equation}

The future state of the tracked object can be  estimated using its dynamics model, applied to each arrival time and destination.  This is identical to the `predict' step of the Kalman filter in \Algo{1} for the future time instant of  interest $t^*>t_n$. For a given arrival time $T=T_i$ and destination $D=d$, the prediction of the object future state is
Gaussian with mean and covariance specified by, respectively,
\eqn{}
{
\hat Z_{t^*|n}^\inb{d,i} &= R_{t^*}^\inb{d,i} \hat Z_{n}^\inb{d,i}+\tilde{m}_{t^*}^\inb{d,i}\\
\Sigma^\inb{d,i}_{t^*|n} &= R^\inb{d,i}_{t^*}\Sigma^\inb{d,i}_{n}(R^\inb{d,i}_{t^*})' + U^\inb{d,i}_{t^*}.
}
Each of $R^\inb{d,i}_{t^*}$, $m^\inb{d,i}_{t^*}$ and $U^\inb{d,i}_{t^*}$ are calculated in the same way as $R_t$, $m_t$ and $U_t$ in \equatn{RUm} for a time step $h$, corresponding to the required prediction time, i.e. $h=t^*-t_n$.  The predicted distribution of the target state  is a Gaussian mixture with the same component weights $u_{i,d}$ as for state estimate at the current time instant $t_n$  in (\ref{eq:StateEstimate}) such that
\begin{align}
&p(X_{t^*}\mid y_{1:t}) \approx \sum_{i=1}^q\sum_{d\in\mathbb{D}} u_{i,d}\normpdf{X_{t^*}}{P_{t}\hat Z^\inb{d,i}_{t^*|n}}{P_{t}\Sigma_{t^*|n}^\inb{d,i}P_t'}.
\end{align}
 The inference algorithms  for the posteriors of the destination, arrival time, and current and future state of the tracked object can be readily parallelised,
with calculations for each quadrature point and/or nominal destination able to  be run largely independently on an independent processor.   Only the weight normalization step as in  e.g. \eqref{eq:StateEstimate} requires results from all calculations to be available, fitting naturally into e.g. a map-reduce programming paradigm for parallel implementation.
\section{Results}\label{sec:Results}
In this section, we evaluate the performance of the proposed bridging-distributions-based intent inference approach in two application areas, namely  HCI\footnote{Please refer to the attached video demonstrating an intent-aware display operating in real-time on a sample of typical in-car free hand  pointing gestures.} and maritime surveillance\footnote{Please refer to the attached videos demonstrating the destination and track prediction for a vessel approaching a coast with multiple possible ports.}. Maximising the likelihood function $\prod_{j=1}^{J}p(y_{1:T}^{j}\mid D=d, \Omega)$ for a sample of $J$\ typical trajectories (constituting the  training set) is the criterion adopted below to set the motion model parameters $\Omega$ . For example for the MRD\ and ERV models, we have $ \Omega=\left\{\Lambda, A , \sigma\right\}$.  The learnt values are then applied to  out-of-sample tracks. This parameter estimation procedure is suitable for an operational real-time system where parameter training is an off-line process  based on historical data.

\subsection{Intent-aware Interactive Displays}
\begin{figure}[t]
\centering
\includegraphics[width=1\linewidth]{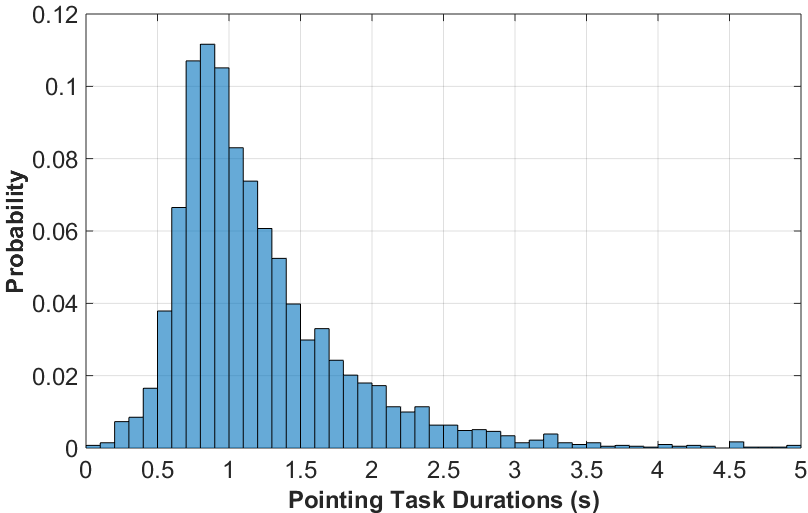}
\caption{\small Distribution of the pointing times, $p(T|D)$, from over 4000 in-vehicle free hand pointing gestures to select  on-screen icons.}
 \label{fig:PointingTime}
 \end{figure}

In the  results presented here, 50 free hand pointing gestures pertaining to four participants  interacting with an in-vehicle  touchscreen   are used (5 tracks are utilised for training). This data is collected in a system mounted to the dashboard of  an instrumented car (identical to the prototype used in \cite{ahmadCyberTrans2015}). It consists of an $11^{"}$ Windows tablet and a gesture-tracker, namely the Leap Motion (LM) Controller \cite{LeapMotion}. The LM produces, in real-time, the 3D\ cartesian coordinates of the pointing finger/hand, i.e. $y_{n}=\left[ \hat x_{t_n}~\hat y_{t_n}~\hat z_{t_n}\right]'$, at an average frame rate of $\approx30\text{Hz}$.
An experimental    GUI of a circular layout is displayed on the tablet screen; it has $21$ selectable circular icons that are  $\approx2$ cm apart. Participants are asked to  select a highlighted on-screen GUI item, giving a known ground truth
intention $D^+$. Nevertheless, in all experiments, the predictor is unaware
of the track end time $T$ and the intended destination, when making
 decisions. To demonstrate the possible range of times of arrival at destination, Figure \ref{fig:PointingTime} depicts the distribution of $T$ for 4,000 recorded in-vehicle free hand pointing gestures aimed at selecting on-screen icons. This empirical $p(T\mid D)$ is used as the arrival time prior for all destinations within the bridging-distribution-based predictors. 

\begin{figure}[t]
\centering
\includegraphics[width=1\linewidth]{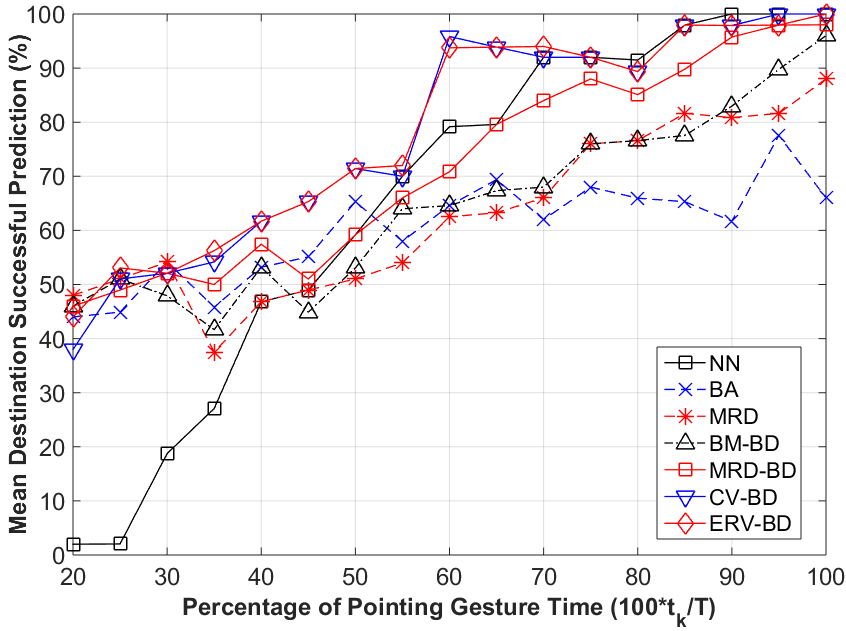}
\caption{\small Mean percentage of successful destination inference as a function of pointing time.}
 \label{fig:Sim_ClassificationSuccess}
 \end{figure}

\begin{figure}[t]
\centering
\includegraphics[width=\linewidth]{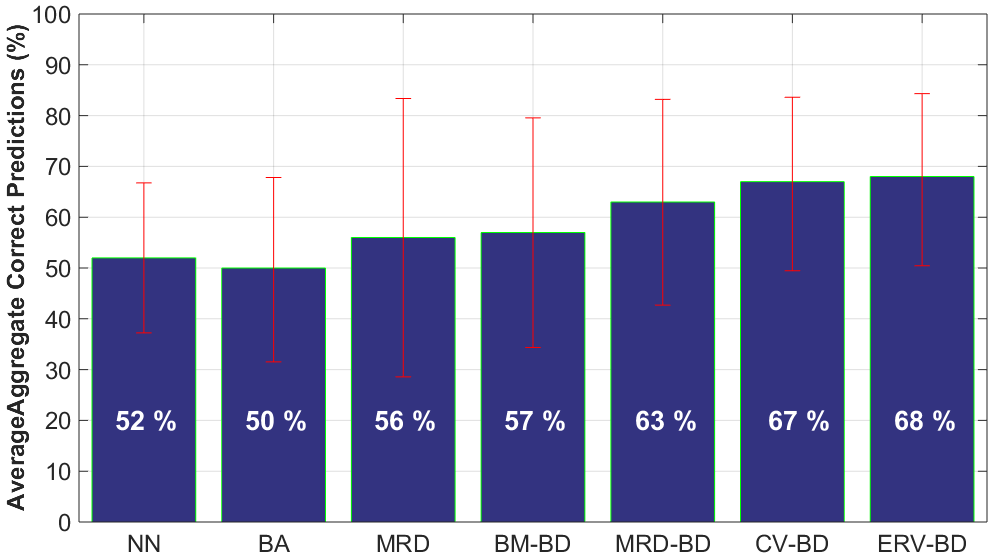}              
\caption{\small Gesture portion (in time) with successful prediction (error bars are one standard deviation).}
\label{fig:Sim_CorrectDecision}
\end{figure}

The inference performance  is evaluated in terms of the ability of the predictor  to  successfully establish the intended icon  $D$ via the MAP\ estimator in (\ref{eq:MAP}), i.e. how early  in the pointing gesture the predictor assigns the highest probability to  $D$. The prediction success is defined  by $S(t_{n})=1 $ if $\hat D(t_{n})=D^+$ and $\mathcal{S}(t_{n})=0 $ otherwise, for observations at times $t_{n}\in \left\{ t_{1},t_{2},...,T\right\}$. This is depicted in  Figure \ref{fig:Sim_ClassificationSuccess} versus the percentage of completed pointing gesture (in time), i.e. $t_{p}=100\times t_{n}/T$, and averaged over all considered pointing tasks. Figure \ref{fig:Sim_CorrectDecision} shows the proportion of the total pointing gesture (in time) for which the  predictor correctly establishes the intended destination. 

In Figures \ref{fig:Sim_ClassificationSuccess} and \ref{fig:Sim_CorrectDecision}, we\ assess the linear destination reverting, BM and  CV models with the bridging prior  notated here as MRD-BD, ERV-BD, BM-BD and CV-BD. A mean reverting diffusion model without bridging  (MRD) is also shown to illustrate the gain attained by incorporating the prior on $X_{T}$. Additionally,   the benchmark Nearest Neighbour (NN) and Bearing Angle (BA)\  methods are examined (see \cite{ahmadCyberTrans2015} for more details). In the former, the destination closest to the pointing finger-tip  position is assigned the highest probability and vice versa; i.e. $p\left(y_{n} |D=d\right)=\mathcal{N}\left({y}_{n};p_{d},{\sigma}_{NN}^{2}\right)$ where ${\sigma}_{NN}^{2}$ is covariance of the multivariate normal distribution. In BA,  $p\left( y_{n} |y_{n-1},D=d\right)=\mathcal{N}\left(\theta_{n} \left\vert 0,{\sigma}_{BA}^{2}\right. \right)$ where $\theta_{n}=\angle\left( y_{n}-y_{n-1},d\right)$ is the angle to destination $D$  and $\sigma_{BA}^{2}$ is a design parameter. It assumes that the cumulative angle to the  intended destination is minimal. 
 
 Figure \ref{fig:Sim_ClassificationSuccess} shows that the introduced bridging-distributions-based inference schemes  CV-BD and ERV-BD, achieve the earliest successful intent predictions. This is particularly visible in the first 70\% of the pointing gesture where notable reductions in the pointing time can be obtained and 
pointing facilitation regimes  can be most  effective (e.g.  expanding icon(s) size, colour, etc.). Destination prediction towards the end of the pointing gesture, e.g. in the last third of the pointing time, has limited benefit, since by that stage the user has already dedicated the necessary visual, cognitive and manual efforts to execute the task.    In general, the performance of all evaluated predictors improves as the pointing hand/finger is closer to the display. This is particulary visible for the nearest neighbour model, which exhibits very poor performance early in the  pointing task and gradually matches other techniques as the observed track length increases  as the pointing finger becomes  close to the endpoint. An exception is the BA model, where the reliability of the heading angle as a measure of intent declines as the pointing finger approaches the destination.

The\ gains from combining the MRD\ motion model with the bridging technique (MRD-BD) are clearly visible in Figure \ref{fig:Sim_ClassificationSuccess} compared to the  MRD without bridging. This can be  attributed to the ability of bridging models to reduce the sensitivity of linear destination reverting models  to  variability in the processed trajectories, which reduces the system's sensitivity to parameter estimates and thus reduces parameter training requirements.  MRD performance, e.g. compared to the NN,\ has notably deteriorated compared to that reported in \cite{ahmadCyberTrans2015} since less  parameter training is performed here; only 5 out of the tested 50 tracks are used for training. Similar observations are made for the ERV which has even more parameters than MRD; the quality of its predictions without bridging (not shown here) is very poor.    

Figure \ref{fig:Sim_CorrectDecision} demonstrates that the proposed bridging-distribution-based  inference  delivers the highest overall correct destination predictions across the pointing trajectories. The highest aggregate successes are achieved by the constant velocity and equilibrium reverting velocity models with relatively tight error bounds. This is due to the importance of the velocity component in the pointing task, which is only captured by these two models. MRD\ without bridging has the largest variance, highlighting its lack of robustness  without the bridging element. NN and BA performances are similar over the considered data set. 

It is important to note that small improvements in pointing task efficiency (effort reduction), even reducing pointing times by
few milliseconds, will have substantial aggregate benefits on
 overall user experience since interactions with displays are very prevalent in typical scenarios, e.g. using a touchscreen in a modern vehicle environment to control the car infotainment system \cite{harvey2013usability,burnett2001ubiquitous}.
\subsection{Maritime Situational Awareness}\label{sec:MSAExperiment}
\begin{figure}[t]
\centering
\includegraphics[width=1\linewidth]{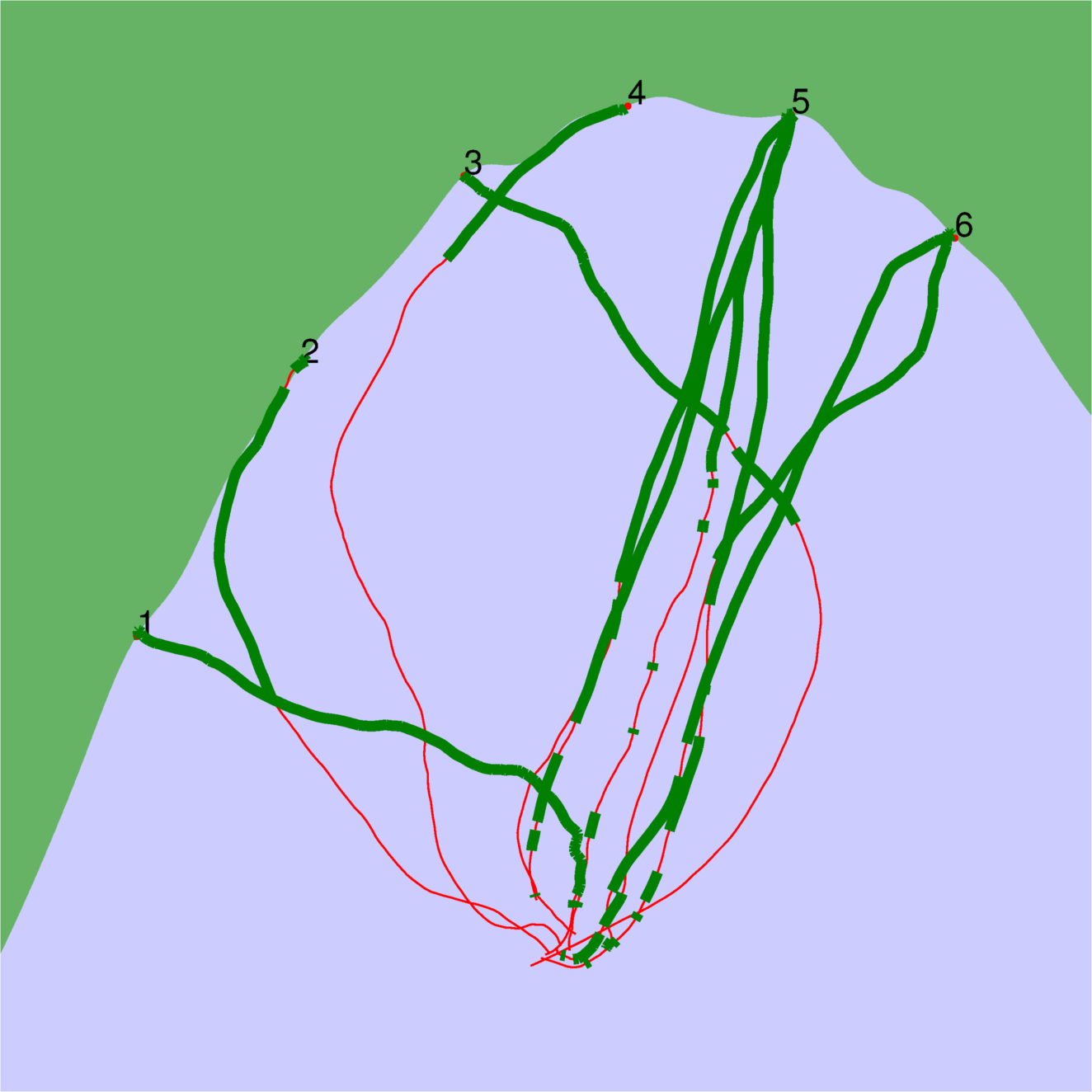}
\caption{\small Destination inference for ten ship trajectories, heading to one of six numbered destinations.  Thick green lines show the portion of each trajectory for which the true intended destination is correctly inferred via MAP estimate in (\ref{eq:MAP}); thin red lines show portions of the trajectory for which this was not the case.}
\label{fig:bayofships}
\end{figure}

\Figure{bayofships} shows the results of the proposed  destination inference algorithm, via the MAP estimate in (\ref{eq:MAP}), for a two-dimensional vessel  tracking problem.  The aim is to predict the endpoint  of each vessel, from a set of $N=6$  possible harbours,  based on noisy observations $y_{1:n}$ of its trajectory up to the present time $t_{n}$. The  trajectory data is generated from a bridged constant velocity model, such that tracks begin at a random point in the middle of the bay (around 20km from the shore). It is conditioned on arriving at a chosen destination port (all equally likely) at a uniformly distributed  random arrival time between 50 and 250 minutes later.  Velocity at arrival is assumed to have a Gaussian distribution with a zero mean and standard deviation 10m min$^{-1}$ (i.e. relatively slow).  The dynamic noise parameter is $\sigma=20\text{m min}^{-3/2}$ in both dimensions.  Observations are  direct measurements of the current vessel position, corrupted by Gaussian noise with standard deviation of 1m. 

Destination inference in Figure \ref{fig:bayofships} is performed using a bridged constant velocity model and the arrival time prior is uniform over the interval [50, 250] minutes, i.e. $p(T\mid D) = \mathcal{U}(50,250)$.  Quadrature was based on  $q=15$ evenly spaced arrival times over this time interval, employing Simpson's rule integration as in (\ref{eq:simpson}).  

For most of the tracks depicted in Figure \ref{fig:bayofships}, the correct intended destination is inferred   early in the observed trajectory.  When the algorithm  fails to achieve such early predictions, e.g.  vessels heading to harbours  2 and 4, visual inspection of these tracks shows that the vessel in question makes a notable sharp manoeuvre at some point in its trajectory. Prior to these sharp changes in direction of travel, the vessel appears to be heading towards a  different endpoint. Nevertheless, after the manoeuvre is completed, the correct destination is quickly inferred. This figure clearly highlights the potential of the proposed bridging-distribution-based destination inference.   
\begin{figure}[t]
\centering
\includegraphics[width=0.8\linewidth]{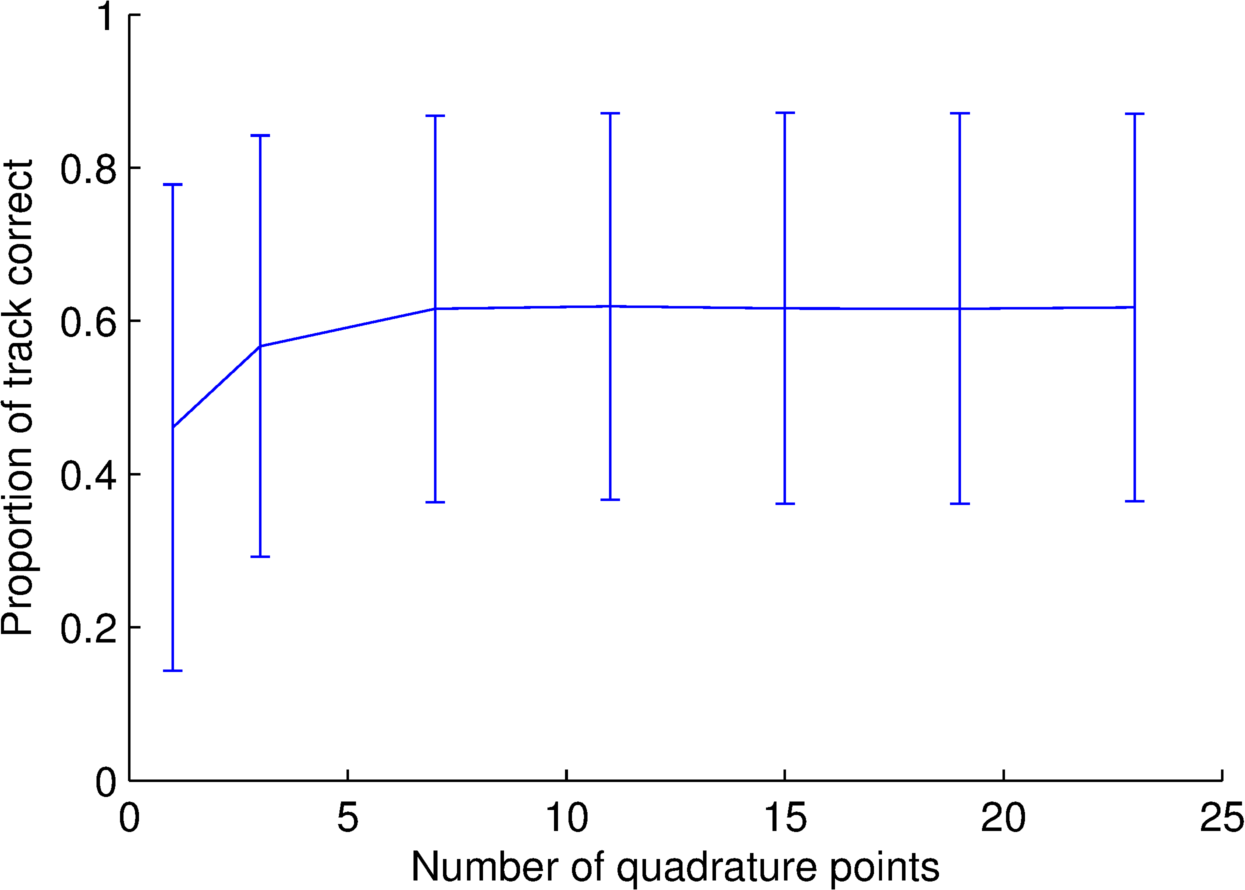}
\caption{\small Proportion of track for which the true destination was inferred for different numbers of quadrature points, using Simpson's rule as in equation (\ref{eq:simpson}), averaged over 100 tracks (error bars show $\pm$ one standard deviation).   End times  are uniformly  distributed in the interval [50, 250], within which the quadrature points are evenly spaced.  For one quadrature point $T=250$ is assumed  for all tracks.}
\label{fig:quadrature}
\end{figure}

\Figure{quadrature} examines the effect of the number of quadrature points $q$  on the endpoint inference outcome for the above vessel tracking scenario.  This  figure shows the average and standard deviation of the aggregate prediction successes for 100 randomly generated vessel trajectories using  Simpson's rule  quadrature as in equation (\ref{eq:simpson}) with varying numbers of quadrature points  $q$. Figure \ref{fig:quadrature}  illustrates that increasing the number of quadrature points (up to $q=9$)  improves the  inference performance, after which the success rate   levels off.  The result for one quadrature point is that for assuming  $T=250$ minutes for all tracks. The results in this figure demonstrate that assuming a distribution of arrival times is beneficial for destination inference compared to guessing a single arrival time. Most importantly, it shows that   only a few quadrature points (e.g. 9 in this case) are sufficient to leverage this benefit. 

\begin{figure}[t]
\centering
\includegraphics[width=0.85\linewidth]{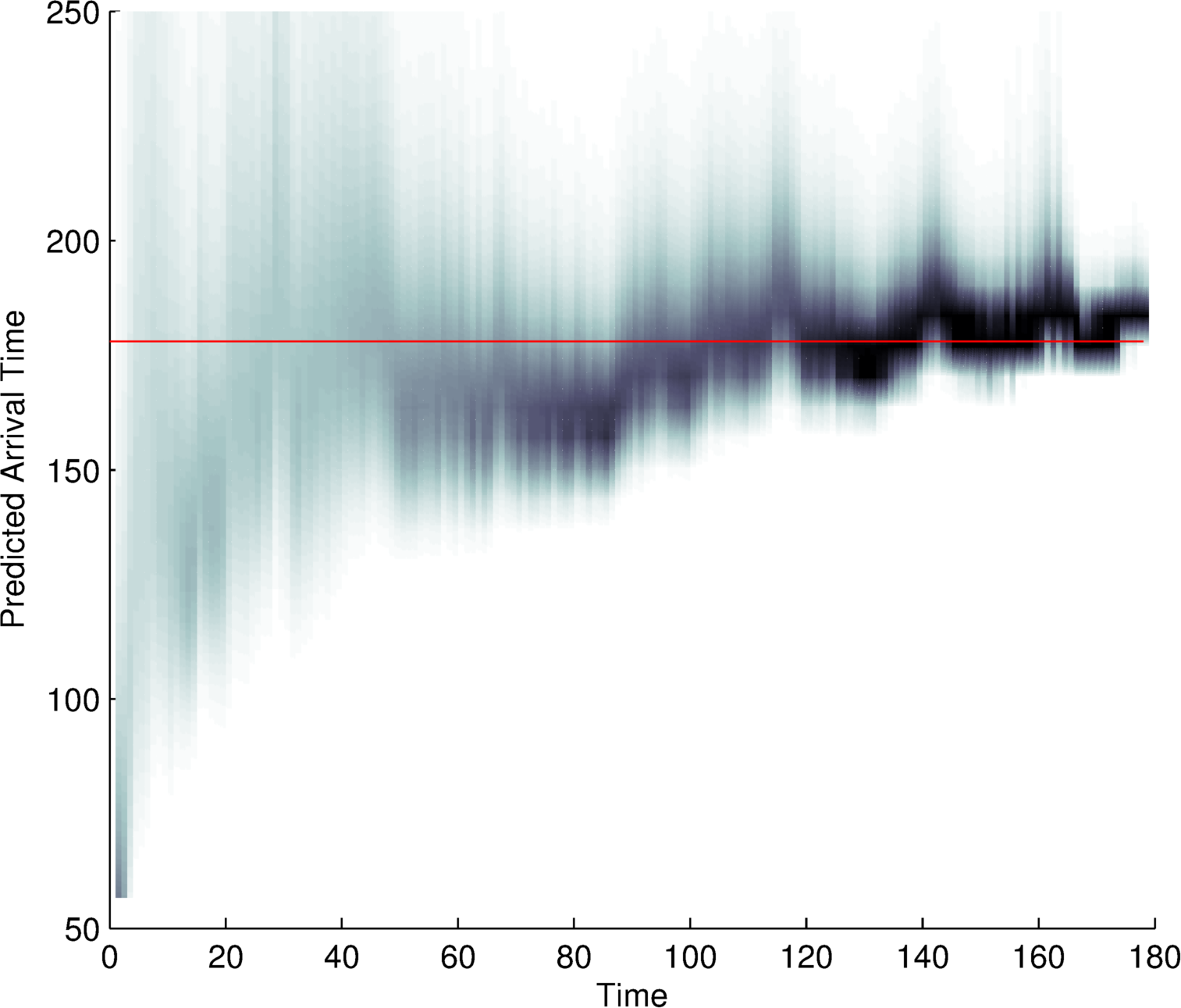}
\caption{\small Arrival time estimate (across all possible destinations) for a track similar to those shown in Figure \ref{fig:bayofships}. The red line shows the true arrival time and shading shows the posterior density at each time.}
\label{fig:arrival}
\end{figure}

Figure \ref{fig:arrival} shows the arrival time estimations for a track similar to those considered in Figure \ref{fig:bayofships}, applying the technique described in Section \ref{sec:arrival-time} with $q=31$ quadrature points.  Initially, the arrival time is uncertain as shown by the diffuse shading. However,  as more trajectory data becomes available, the posterior steadily becomes more concentrated in the region of the true arrival time (red line).

\begin{figure}[t]
\centering
\includegraphics[width=1\linewidth]{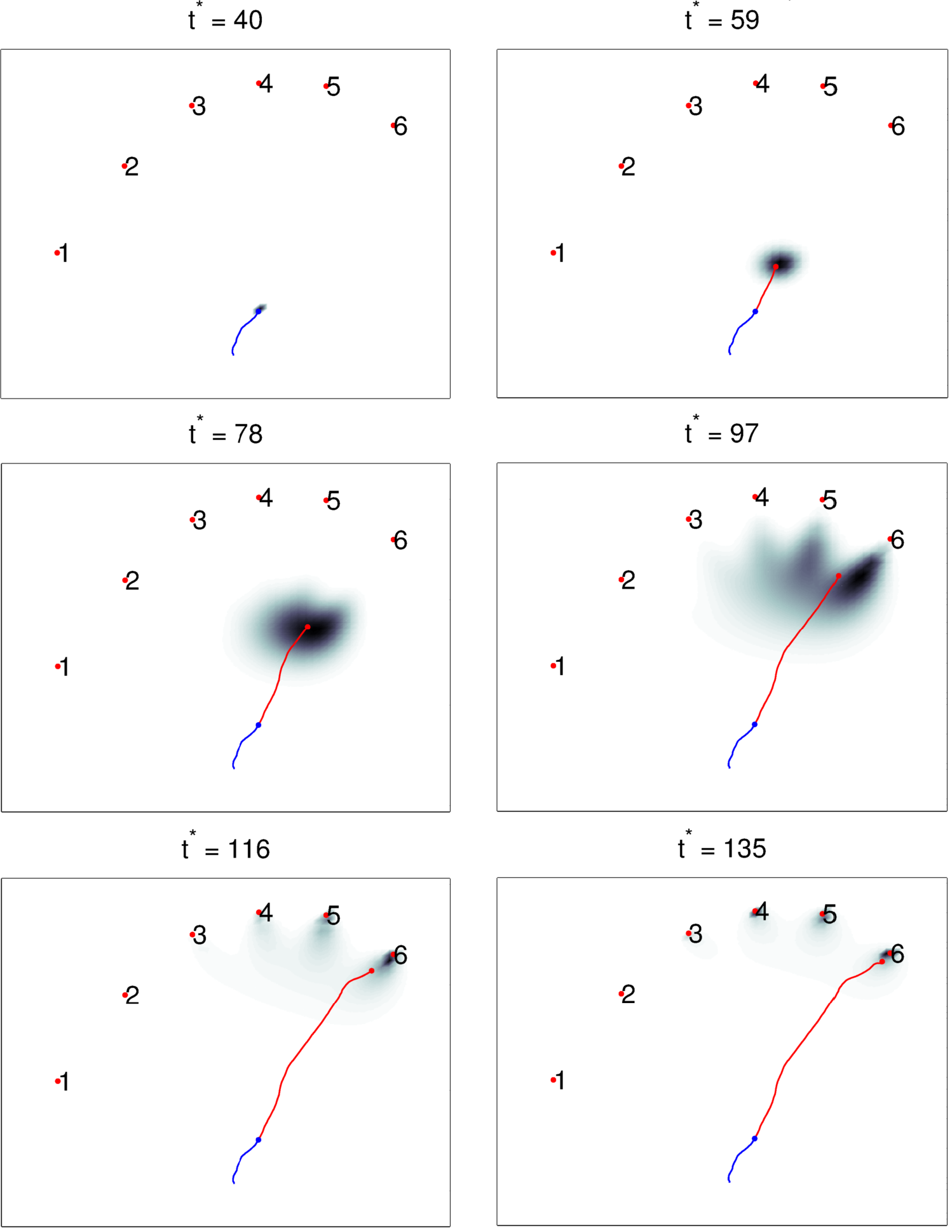}
\caption{\small Predicted target position. Shading shows the posterior distribution of the tracked object position at a range of future time instants $t^*$. The  available trajectory $X_{1:40}$, i.e. at $t_{n}=40$, is depicted by the  solid blue track.  Top left panel shows the current position posterior at $t_{n}=40$ and subsequent panels exhibit the predicted position posterior at times $t^*>40$, up to $t^{*}\approx T$. Red line indicates the  true trajectory values at $t^*$ (red dot is the true target position at $t^*$). The arrival time and destination are unknown.}
\label{fig:predict}
\end{figure}

Finally, the predicted target position at a number of future time instants $t^*>t_n$ are displayed in Figure \ref{fig:predict}, using the prediction method in \ssection{state-prediction}. At the  time shown $t_n$,   40 observations have been made. The tracked object's true position  up to each of the assessed future times  is  shown in red.   Initially, the inference is unimodal, dominated by the object's current motion.  As predictions are made further into the future, the possible destinations  of the target become visibly influential and the predicted position becomes multimodal (e.g. see the last row in Figure \ref{fig:predict}). Each of these modes corresponds to a destination $d\in \mathbb{D}$ and the prediction is dominated by the   endpoint(s) deemed to be more probable by the inference algorithm. For example, the available trajectory  (blue line) leads to more weight being  assigned to endpoints 4, 5 and 6 compared to the other possible destinations (noting that $d=6$ is the true intended destination).  

The effect of the unknown arrival time $T$ (estimated   using $q=25$ quadrature points)   is reflected in Figure \ref{fig:predict}   by  the shape of the predicted densities, which resemble a ``finger'' shape  pointing to each destination (particularly in the fourth panel at $t^*=97$).  These `fingers' result from an increased uncertainty about where the target is in its approach to each destination due to the unknown arrival time. As more observations are made, this effect  diminishes due to the increased confidence in the arrival time, as per Figure \ref{fig:arrival}. 

Defining a normal  trajectory template (pattern of life)  for vessels moving towards potential docking points (e.g. over relatively short distances as in Section \ref{sec:MSAExperiment}) or protected assets can be  challenging. This is due to the fact that the tracked object approach to the intended endpoint can significantly vary and might not necessarily conform to a known pattern (especially if the intent is malicious). Besides, building a complete training data set might not be possible for such scenarios. The proposed inference algorithms in this paper, which do not treat the intent inference problem as an anomaly detection problem, are  able to handle such a setting effectively with  consistent predictions (destination, position and time of arrival) through the use of bridging distributions.      
\section{Conclusion} \label{sec:Conclusions}
This paper sets out a  probabilistic framework for a simple, low complexity  intent inference that demands minimal training and is amenable to parallelisation. Utilising the bridging approach presented here not only permits earlier predictions, but also significantly improves the robustness of destination reverting models against variability in the tracked object behaviour. 
The early inference of the destination, future position and time of arrival of the tracked object, e.g. pointing finger or vessel, can  bring notable benefits such as  reducing the  attention required to   interact with in-vehicle displays, enhancing the ability of maritime surveillance systems and facilitating supervised or unsupervised automated warning/assistive functions. The inference results from the two applications considered in this paper testify to the effectiveness and usefulness of the introduced prediction algorithms.  Whilst linear Gaussian motion and observation models are considered here, extending the formulation  to a more generic settings (nonlinear and/or non-Gaussian),  could broaden its applicability to other scenarios, especially when  targets undertake sharp manoeuvres prior to reaching their intended endpoint. This would require the use of Bayesian filtering techniques such as sequential Monte Carlo algorithms \cite{godsill2007models}, which will entail substantial additional computational cost.
 \section*{Acknowledgment}
 The authors would like to thank Jaguar Land Rover (the Centre for Advanced Photonics and Electronics, CAPE) and the UK Engineering and Physical Science Research Council (BTaRoT grant EP/K020153/1) for funding this research.
\bibliographystyle{IEEEtran}
\bibliography{PredictionReferences}

\end{document}

%% file: jamesmacros.tex
\usepackage[UKenglish]{babel}
\usepackage{amssymb, amsmath, amsthm}
\usepackage{graphicx}
\usepackage[usenames,dvipsnames]{color}
\usepackage{nicefrac}
\usepackage{prettyref}
\usepackage{flushend}
\usepackage{color}
\usepackage{ifthen}
\usepackage{algorithm}
\usepackage[]{algorithmic}
\usepackage{enumitem}
\usepackage{lipsum}

\newcommand{\normdist}[2]{\mathcal{N} \brackets{{#1,#2} }    }
\newcommand{\normpdf}[3]{\mathcal{N} \brackets{{#1; #2, #3} }    }

\newcommand{\diracat}[1]{{\delta_{\left\{ #1 \right\}}}}

\newcommand{\inb}[1]{{(#1)}} 

\newcommand{\thh}{\textsuperscript{th}\ }

\newcommand{\beq}{\begin{eqnarray}}
\newcommand{\eeq}{\end{eqnarray}}
\newcommand{\inmat}[1]{\begin{bmatrix}#1\end{bmatrix}}
\newcommand{\eqn}[2]{\begin{align}\ifthenelse{\equal{#1}{}}{\nonumber}{} #2\ifthenelse{\equal{#1}{}}{\nonumber}{\label{eq:#1}}\end{align}}
\newcommand{\brackets}[1]{\left({#1}\right)}

\newcommand{\equatn}[1]{equation \eqref{eq:#1}}
\newcommand{\equations}[2]{equations \eqref{eq:#1} and \eqref{eq:#2}}

\newcommand{\Equations}[2]{Equations \eqref{eq:#1} and \eqref{eq:#2}}

\newcommand{\fig}[1]{Figure \ref{fig:#1}}

\newcommand{\Figure}[1]{Figure \ref{fig:#1}}
\newcommand{\ssection}[1]{Section \ref{sec:#1}}

\newcommand{\ssections}[2]{Sections \ref{sec:#1} and \ref{sec:#2}}

\newcommand{\Algo}[1]{Algorithm \ref{alg:#1}}

